\documentclass[12pt,a4]{article}
\usepackage{relsize}
\usepackage{array}
\usepackage{a4wide}
\usepackage[hidelinks]{hyperref}
\usepackage{latexsym}
\usepackage{cite}
\usepackage{mathrsfs}
\usepackage{amssymb}
\usepackage{amsmath}
\usepackage{graphicx}
\usepackage{enumerate}
\usepackage[usenames,dvipsnames]{xcolor}
\usepackage{todonotes}
\usepackage{a4wide}
\usepackage{tensor}

\newcommand{\CP}{\mathbb{CP}}

\newcommand{\tr}{\mathrm{tr}}

\newcommand{\be}{\begin{equation}\label}
\newcommand{\ee}{\end{equation}}
\newcommand{\bea}{\begin{eqnarray}\label}
\newcommand{\eea}{\end{eqnarray}}

\newcommand{\und}[1]{\underline{#1}}

\makeatletter
\newcommand*{\textoverline}[1]{$\overline{\hbox{#1}}\m@th$}
\newcommand*\bigcdot{\mathpalette\bigcdot@{.65}}
\newcommand*\bigcdot@[2]{\mathbin{\vcenter{\hbox{\scalebox{#2}{$\m@th#1\bullet$}}}}}
\makeatother

\newcommand{\om}{R}
\newcommand{\omsq}{R^{-2}}
\newcommand{\ii}{\mathrm{i}}

\numberwithin{equation}{section}       

\date{}
\begin{document}

\renewcommand{\thefootnote}{\fnsymbol{footnote}}

   \vspace{1.8truecm}

 \centerline{\LARGE \bf {\sc  Bosonic Symmetries of  }}
  \vskip 12pt
 
 \centerline{\LARGE \bf {\sc  $(2,0)$ DLCQ Field Theories} }
 
\vskip 2cm
  \centerline{
   {\large {\bf  {\sc N.~Lambert,${}^{\,a}$}}\footnote{E-mail address: \href{neil.lambert@kcl.ac.uk}{\tt neil.lambert@kcl.ac.uk}}     \,{\sc A.~Lipstein,$^{\,b}$}\footnote{E-mail address: \href{mailto:arthur.lipstein@durham.ac.uk}{\tt arthur.lipstein@durham.ac.uk}}\, {\sc R.~Mouland${}^{\,a}$}\footnote{E-mail address: \href{rishi.mouland@kcl.ac.uk}{\tt rishi.mouland@kcl.ac.uk}}   {\sc    and P.~Richmond${}^{\,a}$}}\footnote{E-mail address: \href{mailto:paul.richmond@kcl.ac.uk}{\tt paul.richmond@kcl.ac.uk}}  }  
     
\vspace{1cm}
\centerline{${}^a${\it Department of Mathematics}}
\centerline{{\it King's College London }} 
\centerline{{\it The Strand, WC2R 2LS, UK}} 
  
\vspace{1cm}
\centerline{${}^b${\it Department of Mathematical Sciences}}
\centerline{{\it Durham University}} 
\centerline{{\it  Durham, DH1 3LE, UK}}

\begin{abstract}

\noindent We investigate symmetries of the six-dimensional $(2,0)$ theory reduced along a compact null direction. The action for this theory was deduced by considering M-theory on $AdS_7 \times S^4$ and reducing the $AdS_7$ factor along a time-like Hopf fibration which breaks one quarter of the supersymmetry and reduces the isometry group from $SO(6,2)$ to $SU(3,1)$. The boundary theory was previously shown to have 24 supercharges and a Lifshitz scaling symmetry. In this paper, we show that it has four boost-like symmetries and an additional conformal symmetry which furnish a representation of $SU(3,1)$ when combined with the other bosonic symmetries, providing a nontrivial check of the holographic correspondence. 

\end{abstract}

\pagebreak

\tableofcontents

\allowdisplaybreaks

\section{Introduction }\label{Introduction}

One of the most challenging and important open questions in string theory is to describe its strong coupling limit, known as M-theory, whose basic degrees of freedom are not strings but higher dimensional objects called M2 and M5-branes. At low energies, they should be described by superconformal field theories with three-dimensional $\mathcal{N}=8$ and six-dimensional $(2,0)$ supersymmetry, respectively. The lagrangian for an arbitrary number of M2-branes, known as the ABJM theory, turns out to be a superconformal Chern-Simons theory with $\mathcal{N}=6$ supersymmetry which becomes enhanced to maximal supersymmetry quantum mechanically \cite{Aharony:2008ug}. Although an interacting lagrangian with six-dimensional $(2,0)$ supersymmetry does not appear to exist, progress has been made by reducing to five dimensions and interpreting the Kaluza-Klein modes as solitons. Indeed, one of the earliest proposals, known as DLCQ, is to describe the dynamics of M5-branes via quantum mechanics on the moduli space of instantons associated with Kaluza-Klein modes of a null direction\cite{Aharony:1997an}. Moreover, dimensionally reducing the six-dimensional $(2,0)$ theory along a spacelike or timelike direction gives rise to maximal five-dimensional super-Yang-Mills, which was conjectured to provide a complete description of the six-dimensional $(2,0)$ theory nonperturbatively \cite{Douglas:2010iu,Lambert:2010iw,Hull:2014cxa}. Null reductions were subsequently explored in \cite{Lambert:2011gb} and shown to provide a field theory description of the DLCQ proposal \cite{Lambert:2018lgt,Mouland:2019zjr}.

It was recently shown that rescaling a supersymmetric field theory in a way that breaks Lorentz invariance can induce a classical RG flow whose fixed point has enhanced superconformal symmetry. When applied to five-dimensional super-Yang-Mills, this gives rise to a five-dimensional superconformal theory with 24 supercharges, which corresponds to null reduction of the $(2,0)$ theory \cite{Lambert:2019nti}. It was then shown that this mechanism has a natural holographic realisation \cite{Lambert:2019jwi}. The basic idea is to consider M-theory on $AdS_7 \times S^4$ and then write the $AdS_7$ factor as a timelike fibration of a non-compact complex projective space $\tilde{\CP}^3 $, which breaks one quarter of the supersymmetry and the isometry group from $SO(6,2)$ to $SU(3,1)$ \cite{Pope:1999xg}. Flowing to the boundary then gives an $\Omega$-deformed null reduction of the $(2,0)$ theory with 24 superconformal symmetries and a Lifshitz scaling symmetry. 

In this paper, we examine the bosonic symmetries of this theory and find that it has additional boost-like and conformal symmetries which generate an $SU(3,1)$ group when combined with the other bosonic symmetries, as expected from holography. In the $\Omega_{ij} \rightarrow 0$ limit, the theory reduces to the null reduction previously considered in \cite{Lambert:2011gb,Lambert:2019nti} and gains two additional rotational symmetries. There is also a topological charge associated with translations along the null direction, and we show that the Noether charges associated with the new symmetries take a similar form, {\it i.e.}\ they involve integrals over the topological density weighted by functions linear or quadratic in position. These symmetries act very nontrivially on the fields and we obtain an intuitive derivation of them by lifting the five-dimensional action to a six-dimensional diffeomorphism invariant one, although this is not intended to be an action for the six-dimensional $(2,0)$ theory. 

The rest of this paper is organised as follows. In section \ref{Setup} we will give the details of the field theories we are considering and also the (conformal) Killing vectors of the M-theory background which gives rise to them. In section \ref{M} we will construct new bosonic symmetries for the case where the deformation $\Omega_{ij}=0$, corresponding to a null reduction of Minskowski space. Although the results here follow from the $\Omega_{ij}\to 0$ limit of the later results, we find it instructive to consider them separately. In section \ref{O} we will repeat of analysis for the more involved case of $\Omega_{ij}\ne 0$. Section \ref{Conclusion} contains our conclusions and a discussion. We also include two appendices. In appendix \ref{A1} we give an intuitive derivation of the symmetries found in the main section based on assuming a  six-dimensional diffeomorphism invariant action. In appendix \ref{A2} we explicitly show how the (conformal) Killing vectors we found generate $SU(3,1)$, as expected from the $AdS$-dual geometry. 
 
\section{The Field Theories and Background Geometry}\label{Setup}

The fields  in the theories we  consider depend on four space dimensions $x^i$, $i=1,2,3,4$ and a   coordinate $x^-$. Although $x^-$ originates as   a null direction $ (x^0-x^5)/\sqrt2$ in eleven dimensions it plays the role `time' in the field theory. The  action is
\begin{align} \label{somega}
S_\Omega & =     \int dx^-d^4x \, {\cal L}_\Omega\nonumber\\
  &=    \frac{1}{4\pi^2 \om} {\rm tr} \int dx^-d^4x \, \bigg\{\frac12 F_{-i}F_{-i} - \frac12 \nabla_iX^I\nabla_iX^I + \frac12 {\cal F}_{ij}G_{ij}
	 \nonumber\\
&\qquad\qquad \qquad\qquad\qquad -\frac{\ii}{2}\bar\Psi\Gamma_+D_-\Psi + \frac{\ii}{2}\bar\Psi\Gamma_i\nabla_i\Psi - \frac{1}{2}\bar\Psi\Gamma_+\Gamma^I[X^I,\Psi]\bigg\}\ ,
\end{align}
where $I=6,...,10$ is an R-symmetry index labeling five scalars, $G_{ij}$ is a self-dual Lagrange multiplier field and the fermions are real 32-component spinors of $Spin(1,10)$ subject to the constraint $\Gamma_{012345}\Psi=-\Psi$ and $\Gamma_\pm = (\Gamma_0\pm\Gamma_5)/\sqrt2$. Here $\Omega_{ij}=-\Omega_{ji}$ is a constant anti-self-dual two-form with $\Omega_{ik}\Omega_{jk}=\omsq\delta_{ij}$.   We have also introduced
\begin{align}
\nabla_i &= D_i -\frac12\Omega_{ij}x^jD_-\nonumber\\	
{\cal F}_{ij} & = F_{ij} - \frac12\Omega_{ik}x^kF_{-j}+ \frac12 \Omega_{jk}x^kF_{-i}\ ,
\end{align}
with $F_{ij} = \partial_iA_j - \partial_jA_i - \ii[A_i,A_j]$ and $F_{-i} = \partial_-A_i - \partial_iA_- - \ii[A_-,A_i]$. We can also take the special case where $\Omega_{ij}=0$ to obtain
\begin{align}
S_M & =    \int dx^-d^4x \, {\cal L}_M\nonumber\\
 &=   \frac{1}{4\pi^2 \om} {\rm tr} \int dx^-d^4x \, \bigg\{\frac12 F_{-i}F_{-i} - \frac12 D_iX^ID_iX^I + \frac12 F_{ij}G_{ij}
	 \nonumber\\
&\qquad\qquad \qquad\qquad\qquad-\frac{\ii}{2}\bar\Psi\Gamma_+D_-\Psi + \frac{\ii}{2}\bar\Psi\Gamma_iD_i\Psi - \frac{1}{2}\bar\Psi\Gamma_+\Gamma^I[X^I,\Psi]\bigg\}\ .
\end{align}

These field theories arise, after dimensional reduction along $x^+$, from M5-branes on a spacetime whose metric is  \cite{Lambert:2018lgt,Lambert:2019jwi}\footnote{They can also be obtained from a non-Lorentzian rescaling of Yang-Mills gauge theory \cite{Lambert:2019nti}.}
\begin{equation}\label{gdef}
ds^2  \ = \  -2  dx^+ \left(dx^--\frac12 \Omega_{ij}x^idx^j\right)  +  dx^i  dx^i  \ .
\end{equation}
The motivation for considering this metric comes from considering M-theory on $AdS_7 \times S^4$, and writing $AdS_7$ as a timelike circular fibration over a non-compact complex projective space $\tilde{\CP}^3$ \cite{Pope:1999xg}. Restricting to constant $\tilde{\CP}^3$ radius and taking it to infinity then results in \eqref{gdef}, where $\om$ corresponds to twice the $AdS$ radius. Reducing along the fibre breaks one quarter of the supersymmetry, so we expect the boundary theory to have $24$ supercharges. In \cite{Lambert:2019jwi} it was shown that this is indeed the case for $S_\Omega$, which is invariant under 8 supersymmetries and 16 superconformal supersymmetries. Moreover $S_M$ enjoys 16 supersymmetries and 8 superconformal supersymmetries \cite{Lambert:2018lgt,Lambert:2019nti}. For this, and other reasons that will be clear below, we find it instructive to treat $S_M$ separately, even though it formally arises as a special case of $S_\Omega$   when $\Omega_{ij}=0$. 

Here we wish to examine the bosonic symmetries of $S_\Omega$  and $S_M$. It is clear that these actions are invariant under translations in $x^-$ and the 4 rotations of the $x^i$ coordinates which preserve $\Omega_{ij}$ ($S_M$ is invariant under all 6 rotations of $x^i$). Furthermore a little thought shows that they are also invariant under translations in $x^i$, provided that one also shifts $x^-$:
\begin{equation}
	x^i\to x^i + c^i \, , \qquad x^- \to x^-+\frac12\Omega_{ij}c^i x^j\ .
\end{equation}
In each of these cases the fields transform as one would expect under translations and rotations. In addition there is a Lifshitz-type scaling symmetry:
\begin{align}
	x^-\to \lambda x^- \, , \qquad x^i \to \lambda^{\frac12}x^i\ ,
\end{align}
where the fields transform as
\begin{align}\label{weights}
	X^I&\to \lambda^{-1} X^I \, , \qquad \Psi_+ \to \lambda^{-\frac32}\Psi_+ \, , \qquad \Psi_- \to \lambda^{-1}\Psi_-\nonumber\\
	A_-&\to \lambda^{-1} A_- \, ,\qquad\  A_i\to \lambda^{-\frac12}A_i \, , \qquad  \ G_{ij}\to \lambda^{-2} G_{ij}\ ,
\end{align}	
 and $\Psi_{\pm}=-\frac{1}{\sqrt{2}}\Gamma_{\pm}\Gamma_{0}\Psi$. Note that the bosonic symmetries described above form a closed subgroup. Commuting translations along $x^1$ and $x^3$ gives an $x^-$ translation (this is also the case when commuting $x^2$ and $x^4$ translations), but otherwise we obtain the usual algebra of translations, rotations, and a scaling symmetry. 

The large number of supersymmetries suggests that there will be additional bosonic symmetries which, although manifest, are less obvious.  The aim of this work is to find them. For example although these actions do not seem to have a  boost-like symmetry we will see that in fact they do. The bosonic symmetries can also be anticipated from holography. In particular, after reducing $AdS_7$ along the timelike fibre, the bulk isometry group is broken form $SO(6,2)$ to $SU(3,1)$. Remarkably, the bosonic symmetries we find indeed furnish a representation of $SU(3,1)$. The translation, rotation and scaling symmetries mentioned above then form a closed subalgebra of $SU(3,1)$.

We expect that the Killing and conformal Killing vectors of \eqref{gdef} lead to symmetries of the M5-brane. Since we do not have  a lagrangian description for a non-abelian theory of  M5-branes in six-dimensions we are forced to  consider cases where none of the fields depend the $x^+$ direction. In this case the dynamics is described by five-dimensional super-Yang-Mills and its variations. Thus we expect that only those symmetries which leave the fields independent of $x^+$ become symmetries of the reduced non-abelian theory $S_\Omega$. 

Since the M5-brane theory is a conformal field theory we are therefore led to look for solutions to the conformal Killing equation with $\partial_+ k^\lambda=0$:
\begin{equation}
{\cal L}_k g_{\mu\nu} =k^\lambda\partial_\lambda g_{\mu\nu}+\partial_\mu k^\lambda g_{\lambda\nu} + \partial_\nu k^\lambda g_{\mu\lambda} = \omega g_{\mu\nu}\ .
\end{equation}
A vector that satisfies this is called Killing if $\omega=0$ and conformally Killing if $\omega \ne0$. We will use the term (conformally) Killing to describe both cases. For the metric (\ref{gdef}) we find
\begin{align}\label{kis}
k^+ & = \frac14 \omega_2 |x|^2 + v_i x^i +b\nonumber\\
k^- & = \frac12 \omega_2(x^-)^2 + x^-\left(\omega_1+\frac12 v_i\Omega_{ij}x^j\right) + c + \frac12 c_i\Omega_{ij}x^j - \frac{\omega_2}{32} \omsq|x|^4 - \frac18\omsq|x|^2x^kv_{k}\nonumber\\
k^i &= -\frac18\omega_2 \Omega_{ki}x^k|x|^2 - \frac12 x^kv_k\Omega_{li}x^l + c_i + M_{ij}x^j + \frac12 \omega_1 x^i + \frac12 v^kx^l\Omega_{kl} x^i - \frac14 |x|^2v^k\Omega_{ki} \nonumber\\ &\qquad + x^- \left(v^i+ \frac12\omega_2 x^i \right)\nonumber\\
\omega & = \omega_1 + v_i\Omega_{ij}x^j + \omega_2 x^-\ ,
\end{align} 
where $b,c,c_i,M_{ij},\omega_1,v_i,\omega_2$ are all constant independent parameters with $M=-M^T$, $[M,\Omega]=0$. This corresponds to $1+1+4+4+1+4+1=16$ (conformal) Killing vectors.  Recall that six-dimensional Minkowski space admits 21 Killing vectors and 7 conformal  Killing vectors. The metric (\ref{gdef}) is conformal to Minkowski space and so must also admit a total of 28 (conformal) Killing vectors. We conclude that 12 must depend on $x^+$. 

In particular taking special cases we have the following types of $x^+$-independent (conformal) Killing vectors:
\begin{align}\label{list}
&{\rm type}\ I \ \ \ \  \left(b,0,0,0,0,0\right)\nonumber\\
&{\rm type}\ II \ \ \ \left(0,c,0,0,0,0\right)\nonumber\\
&{\rm type}\ III\ \   \left(0,\frac12 c_i\Omega_{ij}x^j,c_i\right)\nonumber\\
&{\rm type}\ IV \ \ \   \left(0,0,M_{ij}x^j\right)\nonumber\\
&{\rm type}\ V \ \ \ \   \left(0,\omega_1 x^-,\frac12 \omega_1x_i\right)\nonumber\\
&{\rm type}\ VI \  \left(v_ix^i,\frac12 x^-v_i\Omega_{ij}x^j - \frac18\omsq|x|^2x^kv_k,x^-v_i - \frac12 x^kv_k\Omega_{li}x^l +\frac12 v^kx^l\Omega_{kl}x^i-\frac14|x|^2v^k\Omega_{ki}\right)\nonumber\\
&{\rm type}\ 
VII \    \left(\frac14 \omega_2|x|^2,\frac12\omega_2(x^-)^2- \frac{\omega_2}{32} \omsq|x|^4,-\frac{1}{8}\omega_2\Omega_{ki}|x|^2x_k+ \frac12\omega_2x_ix^-\right)\ .
\end{align}

The type $I$ symmetry is a translation in $x^+$ and acts trivially in the five-dimensional lagrangian. Nevertheless we identify the associated conserved current with the topological current
\begin{align}
P_{+}^-  = \frac{1}{32\pi^2 \om}\varepsilon^{ijkl}\tr(F_{ij}F_{kl}) \, , \qquad 	P_{ + }^i = -\frac{1}{8\pi^2 \om}\varepsilon^{ijkl}\tr(F_{-j}F_{kl})\ .
\end{align} 
In particular the conserved charge, corresponding to momentum along $x^+$, is identified with the instanton number of the gauge field on ${\mathbb R}^4$
\begin{align}
p_+= \int d^4x \, P^-_{+} =  \frac{1}{32\pi^2 \om}\varepsilon^{ijkl}\int d^4x \, \tr(F_{ij}F_{kl})	\ .
\end{align}
Next we observe that type $II$, $III$ and $IV$ are Killing vectors with $\omega=0$. Type $II$ and $III$ are the 5 remaining translations whereas type $IV$ are the 4 rotations that preserve $\Omega$.  Type $V$ is the Lifshitz scaling symmetry with $\omega\ne 0$. Thus the  symmetries corresponding to types $I$-$V$ are easy to identify. However type $VI$ and type $VII$ are   new   non-trivial bosonic symmetries. In appendix \ref{A2}, we derive the conformal Killing vectors in \eqref{list} from the Killing vectors of $AdS_7$ reduced along a timelike fibre. This construction implies that the underlying symmetry group is $SU(3,1)$, which can be explicitly verified by computing Lie derivatives of the conformal Killing vectors. One can also check that the Killing vectors generated by type $I$-$V$ close among themselves to form a subalgebra.

\section{Minkowski Space: $\Omega_{ij}=0$}\label{M}

Before we address the bosonic symmetries of $S_\Omega$ it is worthwhile to first find the symmetries of  $S_M$. In particular $S_M$ arises from  dimensional reduction  along $x^+$ of Minkowski space  in the lightcone coordinates obtained from (\ref{gdef}) (with $\Omega_{ij}=0$).  If we
simply set $\Omega_{ij}=0$ in (\ref{list}) we find the following (conformal) Killing vectors of Minkowski space
\begin{align}\label{list0}
{\rm type}\ I \qquad  &\left(b,0,0,0,0,0\right)\nonumber\\
{\rm type}\ II \qquad & \left(0,c,0,0,0,0\right)\nonumber\\
{\rm type}\ III\qquad   &\left(0,0,c_i\right)\nonumber\\
{\rm type}\ IV \qquad   &\left(0,0,M_{ij}x^j\right)\nonumber\\
{\rm type}\ V \qquad  & \left(0,\omega_1 x^-,\frac12 \omega_1x_i\right)\nonumber\\
{\rm type}\ VI \qquad & \left(v_ix^i,0,x^-v_i \right)\nonumber\\
{\rm type}\ VII \qquad &  \left(\frac14 \omega_2|x|^2,\frac12\omega_2(x^-)^2,\frac12\omega_2 x^-x_i\right)\ .
\end{align}
These should all lead to symmetries of $S_M$. As before the first five types are simply translations, rotations and a Lifshitz scaling.    Note that now there are 6 rotations since the constraint $[M,\Omega]=0$ is vacuous. Hence we find 16 Killing vectors and 2 conformal Killing vectors that do not depend on $x^+$. The associated generators form a subalgebra of the six-dimensional conformal algebra that commute with $P_+$ and were discussed in \cite{Aharony:1997an} within the context of a DLCQ description of M5-branes.

As an aside we note that Minkowski space has 21 Killing and 7 conformally Killing vectors. Therefore it follows from our derivation that the  additional 10 (conformal) Killing vectors which are not in (\ref{list0}) must  depend on $x^+$.  Thus 2 of the 12 $x^+$ dependent (conformal) Killing vectors for $\Omega_{ij}\ne 0$   become the additional $x^+$-independent rotations   when $\Omega_{ij}=0$ while the other 10 remain $x^+$ dependent. Of these 5 are  Killing vectors of Minkowski space corresponding boosts in the $(x^+,x^-)$ and $(x^+,x^i)$-planes:
\begin{align}
{\rm type}\ VIII \qquad  &\left(ax^+,-ax^-,0,0,0,0\right)\nonumber\\
	{\rm type}\ IX \qquad &\left(0,u_ix^i,x^+u_i \right)\ .
\end{align}
The remaining   missing 5 conformal Killing vectors are then found to be:
\begin{align}
{\rm type}\ X  \qquad  & \left(-2 w_i x^i x^+, -2w_i x^i x^-, |x|^2 w_i - 2 x^+ x^- w_i - 2 w_j x^j x^i\right)\nonumber\\
	{\rm type}\  XI \qquad &\left(\frac12\omega_4(x^+)^2,\frac14 \omega_4|x|^2,\frac12\omega_4 x^+x_i\right)\ ,
\end{align}
with $\omega = -4 w_i x^i + \omega_4 x^+$.
However we are not interested in any of these as they depend on $x^+$ and hence cannot lead to symmetries of the five-dimensional action. What remains is to show that types $VI$ and $VII$ lead to symmetries of the five-dimensional non-abelian theory. 

\subsection{Type $VI$}

Let us look at type $VI$. This is a Killing vector  and corresponds to the six-dimensional diffeomorphism
\begin{equation}\label{rot}
x^+\to x^+ +v^ix^i \, ,\qquad x^-\to x^- \, ,\qquad x^i \to x^i + v^i x^-\ .
\end{equation}
These can be thought of as null boosts and in particular they are a part of the six-dimensional Lorentz group.  A traditional boost consists of combining a left-moving and a right-moving null boost.
To continue we assume that all the fields are independent 
of $x^+$. In  this case we find the variations of a Galilean boost in five-dimensions: 
\begin{align}
	\delta_{} x^- = 0\, , \qquad \delta_{} x^i = v^ix^-\ .
\end{align}  
In addition to this transformation  we postulate a further tensor-like variation
\begin{align}\label{Achange}
\delta_{} A_- &= -v^iA_i\nonumber\\
\delta_{}A_i &=0\nonumber\\
\delta_{}X^I&=0\nonumber\\
\delta_{} G_{ij} &= -2\bigg(F_{-[i}v_{j]} + \frac12\varepsilon_{ijkl}F_{-k}v_l\bigg)\nonumber\\
\delta_{}\Psi  &= \frac{1}{2}v^j\Gamma_+\Gamma_j\Psi \ .
\end{align}
An intuitive, six-dimensional,  derivation of these expressions is given in appendix \ref{A1}. We find
\begin{align}
\delta_{}S_{M} 
&= \frac{1}{4\pi^2\om}\int dx^-d^4x \,
\bigg\{\frac12\varepsilon_{ijkl}{\rm tr}( F_{ij}F_{-l})v_k \bigg\}
\ .
\end{align}
This term can be identified with  $\frac12{\rm tr}(F\wedge F)\wedge v$ and hence is a total derivative.
 It is interesting to note that, even in the abelian case, the action is only invariant under the action of the six-dimensional Lorentz group up to the boundary term $\frac12{\rm tr}(F\wedge F)\wedge v $.

Using the standard formula the associated Noether current takes the somewhat unconventional form (we have set the fermions to zero for simplicity - the full results can be found by setting $\Omega_{ij}=0$ in \eqref{pm} and \eqref{pi}): 
\begin{align}
P^-(v) & = -\frac{1}{4\pi^2\om}x^-{\rm tr}(F_{-i}v^j\partial_jA_i) + \frac{1}{32\pi^2\om}v^mx^m\varepsilon_{ijkl}{\rm tr}(F_{ij}F_{kl})\nonumber\\
P^i(v) &=-\frac{1}{8\pi^2\om}v^mx^m\varepsilon_{ijkl}{\rm tr} ( F_{-j}F_{kl}) + \frac{1}{4\pi^2\om} {\rm tr}(F_{-i} v^jA_j) \nonumber\\
&\ \ \ \ -\frac{1}{4\pi^2\om}x^-{\rm tr}\left(-F_{-i}v^j\partial_jA_-+ G_{ik}v^j\partial_jA_k-D_iX^I v^j\partial_jX^I \right) - x^-v^i{\cal L}_M\ .
\end{align}  
This satisfies $\partial_-P^-(v)+\partial_i P^i (v)=0$ on-shell. Since these are null boosts we associate the conserved charge with a momentum along the $x^i$-direction:
\begin{equation}
	p_i  = \int d^4x \, P^-(v_j=\delta_j^i) = \frac{1}{32\pi^2\om}\int_{x^-=0} d^4x\,    x^i \varepsilon_{klmn}{\rm tr}(F_{kl}F_{mn})\ ,
\end{equation}
where, since $p_i$ is independent of $x^-$, we have simplified the expression by evaluating it at  $x^-=0$.

\subsection{Type $VII$}

Next, let us consider the type $VII$ transformation.   In six-dimensions this is the diffeomorphism
\begin{equation}
x^+\to x^+ +\frac12\omega_2|x|^2\, ,\qquad 
x^-\to x^- + \frac12 \omega_2 (x^-)^2 \, ,\qquad x^i \to x^i + \frac12\omega_2 x^-x^i\ .
\end{equation}
Reducing to five dimensions we find
\begin{align}
	\delta_{} x^- = \frac12\omega_2(x^-)^2\, ,\qquad \delta_{} x^i =  \frac12 \omega_2 x^- x^i  \ .
\end{align}
In this case the measure is rescaled
\begin{align}
	\delta_{}(dx^-d^4x) = 3\omega_2 x^- (dx^-d^4x)\ .
\end{align} 
We find we need a transformation that acts like a six-dimensional tensorial transformation of the fields along with a Lifshitz rescaling:
\begin{align}
\delta_{} A_{-}&= -\omega_2 x^- A_--\frac12\omega_2 x^iA_{i} \nonumber\\
 \delta_{} A_{i}&=-\frac12\omega_2 x^-A_i\nonumber\\ 
 \delta_{} X^I &=- \omega_2 x^- X^I \nonumber\\
 \delta_{} G_{ij}&=  -2\omega_2 x^-G_{ij}-\omega_2 \left(F_{-[i}x_{j]}+\frac{1}{2}\epsilon_{ijkl}F_{-k}x^{l}\right)\nonumber\\
\delta_{}\Psi  &= - \frac14\omega_2(5+\Gamma_{-+}) x^-\Psi+\frac{1}{4}\omega_2 x^j\Gamma_+\Gamma_j\Psi \ .
\end{align}
The action is now invariant, up to a total derivative. Indeed the main difference with the type $VI$ case is to replace $v^i$ with $\tfrac12 \omega_2 x^i$ in the calculations. 
The conserved Noether current now takes the form
\begin{align}
K^-  & = -\frac1{8\pi^2 \om} x^-{\rm tr}(F_{-i}x^j\partial_jA_i) -\frac1{8\pi^2 \om} x^-{\rm tr}(F_{-i} A_i ) + \frac1{128\pi^2 \om}|x|^2\varepsilon_{ijkl}{\rm tr}(F_{ij}F_{kl})    + {\cal O}((x^-)^2)\nonumber\\
K^i  &=-\frac1{32\pi^2 \om}|x|^2\varepsilon_{ijkl}{\rm tr} ( F_{-j}F_{kl}) +\frac1{8\pi^2 \om}  {\rm tr}(F_{-i}x^jA_j)   +{\cal O}(x^-) \ ,
\end{align}
where we have omitted fermions and terms that are higher order in  $x^-$ for simplicity. Again the full results can be found by setting $\Omega_{ij}=0$ in \eqref{km} and \eqref{ki}. As above these are not needed if one evaluates the charge at $x^-=0$. In particular the conserved charge is
\begin{equation}
\mathrm{k} = \int d^4x \, K^-= \frac1{128\pi^2 \om}\int_{x^-=0}d^4x \ |x|^2\varepsilon_{ijkl}{\rm tr}(F_{ij}F_{kl}) \ .
\end{equation}
 
\section{Symmetries for $\Omega_{ij}\ne 0$}\label{O}

For $\Omega_{ij}\ne 0 $ the conformal transformation 
\begin{align}
ds^2 \to \frac{ds^2}{\cos^2(x^+/2\om)}	\ ,
\end{align}
maps the metric (\ref{gdef}) to a flat metric. In particular points with $x^+\in (-\pi \om,\pi \om)$ cover all of six-dimensional Minkowski space. Thus for $\Omega_{ij}\ne0$, $x^+$ naturally lies in a finite range (but need not be periodic) whereas for $\Omega_{ij}=0$ restricting $x^+$ to lie in a finite range in (\ref{gdef}) requires an ad hoc compactification. It was further suggested in \cite{Lambert:2019jwi} that one could double the range of $x^+$ by imposing `reflecting' boundary conditions to make the fields periodic with period $4\pi \om$. In this case we should replace $\om\to \om/2$ in the action. 

As mentioned in section \ref{Setup}, the metric in (\ref{gdef}) arises from constructing $AdS_7$ as a timelike Hopf-fibration over a non-compact complex projective space and going to the boundary\cite{Pope:1999xg}. Moreover, in appendix \ref{A2}, we show that the conformal Killing vectors in \eqref{list} generate $SU(3,1)$, which is the residual isometry group of $AdS_7$ after reducing along the fibre. Verifying that they correspond to symmetries of the action in \eqref{somega} therefore provides a nontrivial check of the holographic correspondence.

In the rest of this section we extend our results above to the general case of $\Omega_{ij}\ne0$. The expressions here are considerably more complicated but their motivation can be found in appendix \ref{A1}. Otherwise the analysis is similar to the $\Omega_{ij}=0$ case above so we will be more succinct in our discussion. 

\subsection{Type $VI$}

In six-dimensions the conformal Killing vector leads to the following diffeomorphism
\begin{align}
x^+&\to x^+ +v_ix^i\nonumber \\ 
x^-&\to x^- + \frac{1}{2} \Omega_{ij} v_i x^j x^- - \frac{1}{8} \omsq |x|^2 v_i x^i \nonumber \\
x^i &\to x^i +\frac{1}{2}\Omega_{jk}v_j x^k x^i + v_i x^- + \frac{1}{2}\Omega_{ij}v_k x^j x^k + \frac{1}{4} |x|^2 \Omega_{ij} v_j
\ ,
\end{align} so that 
upon reduction to five dimensions we find
\begin{align}
\delta_{} x^-&=\frac{1}{2} \Omega_{ij} v_i x^j x^- - \frac{1}{8} \omsq |x|^2 v_i x^i \nonumber\\
 \delta_{} x^i  &= \frac{1}{2}\Omega_{jk}v_j x^k x^i + v_i x^- + \frac{1}{2}\Omega_{ij}v_k x^j x^k + \frac{1}{4} |x|^2 \Omega_{ij} v_j\ .
\end{align}
This time we find that the measure is now rescaled
\begin{align}
	\delta_{}(dx^-d^4x) = 3\Omega_{ij}v_i x^j (dx^-d^4x)\ .
\end{align}Following appendix \ref{A1}  we find  
\begin{align} 
  \delta_{} A_- &= -\frac{1}{2} \Omega_{ij} v_i x^j A_-  - v_i A_i \nonumber\\
  \delta_{} A_i &= -\frac{1}{2} \Omega_{jk} v_j x^k A_i+ \frac{1}{2}\left( \Omega_{ij} v_k x^k + \Omega_{ik} v_k x^j-\Omega_{jk}( v_i x^k + v_k x^i ) \right)A_j\nonumber\\
  &\qquad + \frac{1}{8} \left( \omsq |x|^2 v_i + 2\omsq v_j x^j x^i + 4 \Omega_{ij} v_j x^- \right) A_- \nonumber\\
  \delta_{} X^I &= -\Omega_{ij} v_i x^j X^I  \nonumber\\
    \delta_{} G_{ij} &=  -2\Omega_{kl} v_k x^l G_{ij}  - \frac{1}{2}\left( \lambda^{ki} G_{kj} - \lambda^{kj} G_{ki} +\varepsilon_{ijkl} \lambda^{mk} G_{ml} \right) + v_i F_{-j} - v_j F_{-i} + \varepsilon_{ijkl} v_k F_{-l}\nonumber\\
  \delta_{} \Psi &=-\frac{1}{4}(5+\Gamma_{-+})\Omega_{ij} v_i x^j \Psi+  \frac{1}{2} v_i \Gamma_{+}\Gamma_{i}\Psi + \frac{1}{4} \lambda^{ij}\Gamma_{ij}\Psi\ ,
\end{align}
where 
\begin{align}
\lambda^{ij}=\frac{1}{2}\left( \Omega_{ij} v_k x^k + \Omega_{ik} v_k x^j - \Omega_{jk} v_k x^i +\Omega_{ik} v_j x^k - \Omega_{jk} v_i x^k \right)\ .	
\end{align}

The conserved Noether current now takes the form (where now $i$ labels the four choices for $v^j = \delta^j_i$)
\begin{align} \label{pm}
  \tensor{P}{^-_i} &= \left( \tfrac{1}{2} x^- \Omega_{ij} x^j - \tfrac{1}{8} \omsq |x|^2 x^i \right)\mathcal{L}_\Omega\nonumber\\
  & + \frac{1}{4\pi^2 \om}\text{tr}\,\Bigg[\,\, \tfrac{1}{8} x^i \varepsilon_{jklm} F_{jk} F_{lm} -\tfrac{1}{4} \omsq x^i X^I X^I \nonumber\\
  &\hspace{12mm} + \left( F_{-j} + \tfrac{1}{2} \Omega_{kl} x^l G_{jk} \right) \bigg( -\tfrac{1}{2} \Omega_{im} x^m A_j \nonumber\\
  &\hspace{55mm} + \tfrac{1}{2}\left( \Omega_{jm} x^i - \Omega_{ij} x^m -\delta_{ij} \Omega_{mn} x^n + \Omega_{im}x^j \right) A_m \nonumber\\
  &\hspace{55mm} + \tfrac{1}{8} \left( \delta_{ij} \omsq |x|^2 + 2\omsq x^i x^j - 4\Omega_{ij} x^- \right) A_- \nonumber\\
  &\hspace{55mm} - \left( \tfrac{1}{2} x^- \Omega_{im} x^m - \tfrac{1}{8}\omsq |x|^2 x^i \right) \partial_- A_j \nonumber\\
  &\hspace{55mm} - \left( x^- \delta_{im} + \tfrac{1}{2} \Omega_{mn} x^n x^i + \tfrac{1}{2}\Omega_{in} x^n x^m - \tfrac{1}{4} \Omega_{im} |x|^2 \right) \partial_m A_j \bigg) \nonumber\\
  &\hspace{12mm} -\tfrac{1}{2} \Omega_{jk} x^k \left( \nabla_j X^I \right) \bigg( \Omega_{il} x^l X^I + \left( \tfrac{1}{2} x^- \Omega_{il} x^l - \tfrac{1}{8}\omsq |x|^2 x^i \right)\partial_- X^I \nonumber\\
  &\hspace{49mm} + \left( x^- \delta_{il} + \tfrac{1}{2}\Omega_{lm} x^m x^i + \tfrac{1}{2} \Omega_{im} x^m x^l - \tfrac{1}{4} |x|^2 \Omega_{il} \right) \partial_l X^I \bigg) \nonumber\\
  &\hspace{12mm}-\tfrac{\ii}{2}\bar{\Psi} \left( \Gamma_+ + \tfrac{1}{2} \Omega_{jk} x^k \Gamma_j  \right)\bigg( -\tfrac{1}{4} \Omega_{il} x^l \left( 5+\Gamma_{-+} \right) \Psi + \tfrac{1}{2} \Gamma_+ \Gamma_i \Psi \nonumber\\
  &\hspace{58mm} + \tfrac{1}{8} \left( \Omega_{lm} x^i + \Omega_{li} x^m - \Omega_{mi} x^l + \delta_{mi} \Omega_{ln} x^n - \delta_{li} \Omega_{mn} x^n \right) \Gamma_{lm} \Psi \nonumber\\
  &\hspace{58mm} -\left( \tfrac{1}{2} x^- \Omega_{il} x^l - \tfrac{1}{8}\omsq |x|^2 x^i \right)\partial_-\Psi \nonumber\\
  &\hspace{58mm} -\left( x^- \delta_{il} + \tfrac{1}{2}\Omega_{lm} x^m x^i + \tfrac{1}{2} \Omega_{im} x^m x^l - \tfrac{1}{4} |x|^2 \Omega_{il} \right) \partial_l \Psi \bigg) \Bigg]\ ,
\end{align}
\begin{align} \label{pi}
  \tensor{P}{^j_i} &= \left( x^- \delta_{ij} + \tfrac{1}{2}\Omega_{jk} x^k x^i + \tfrac{1}{2} \Omega_{ik} x^k x^j -\tfrac{1}{4} \Omega_{ij} |x|^2 \right)\mathcal{L}_\Omega\nonumber\\
  & + \frac{1}{4\pi^2 \om}\text{tr}\,\Bigg[\,\, -\tfrac{1}{2} x^i \varepsilon_{jklm} F_{-k} F_{lm} -\tfrac{1}{2} \Omega_{ij}  X^I X^I \nonumber\\
  &\hspace{12mm} + \left( F_{-j} + \tfrac{1}{2} \Omega_{kl} x^l G_{jk} \right) \bigg( \tfrac{1}{2} \Omega_{im} x^m A_- + A_i \nonumber\\
  &\hspace{55mm} + \left( \tfrac{1}{2} x^- \Omega_{im} x^m - \tfrac{1}{8}\omsq |x|^2 x^i \right) \partial_- A_- \nonumber\\
  &\hspace{55mm} + \left( x^- \delta_{im} + \tfrac{1}{2} \Omega_{mn} x^n x^i + \tfrac{1}{2}\Omega_{in} x^n x^m - \tfrac{1}{4} \Omega_{im} |x|^2 \right) \partial_m A_- \bigg) \nonumber\\
  &\hspace{12mm} + G_{jk} \bigg( -\tfrac{1}{2} \Omega_{im} x^m A_k  + \tfrac{1}{2}\left( \Omega_{km} x^i - \Omega_{ik} x^m -\delta_{ik} \Omega_{mn} x^n + \Omega_{im}x^k \right) A_m \nonumber\\
  &\hspace{27mm} + \tfrac{1}{8} \left( \delta_{ik} \omsq |x|^2 + 2\omsq x^i x^k - 4\Omega_{ik} x^- \right) A_- \nonumber\\
  &\hspace{27mm} - \left( \tfrac{1}{2} x^- \Omega_{im} x^m - \tfrac{1}{8}\omsq |x|^2 x^i \right) \partial_- A_k \nonumber\\
  &\hspace{27mm} - \left( x^- \delta_{im} + \tfrac{1}{2} \Omega_{mn} x^n x^i + \tfrac{1}{2}\Omega_{in} x^n x^m - \tfrac{1}{4} \Omega_{im} |x|^2 \right) \partial_m A_k \bigg) \nonumber\\
  &\hspace{12mm} + \left( \nabla_j X^I \right) \bigg( \Omega_{il} x^l X^I + \left( \tfrac{1}{2} x^- \Omega_{il} x^l - \tfrac{1}{8}\omsq |x|^2 x^i \right)\partial_- X^I \nonumber\\
  &\hspace{35mm} + \left( x^- \delta_{il} + \tfrac{1}{2}\Omega_{lm} x^m x^i + \tfrac{1}{2} \Omega_{im} x^m x^l - \tfrac{1}{4} \Omega_{il} |x|^2  \right) \partial_l X^I \bigg) \nonumber\\
  &\hspace{12mm}+\tfrac{\ii}{2}\bar{\Psi} \Gamma_j\bigg( -\tfrac{1}{4} \Omega_{il} x^l \left( 5+\Gamma_{-+} \right) \Psi + \tfrac{1}{2} \Gamma_+ \Gamma_i \Psi \nonumber\\
  &\hspace{30mm} + \tfrac{1}{8} \left( \Omega_{lm} x^i + \Omega_{li} x^m - \Omega_{mi} x^l + \delta_{mi} \Omega_{ln} x^n - \delta_{li} \Omega_{mn} x^n \right) \Gamma_{lm} \Psi \nonumber\\
  &\hspace{30mm} -\left( \tfrac{1}{2} x^- \Omega_{il} x^l - \tfrac{1}{8}\omsq |x|^2 x^i \right)\partial_-\Psi \nonumber\\
  &\hspace{30mm} -\left( x^- \delta_{il} + \tfrac{1}{2}\Omega_{lm} x^m x^i + \tfrac{1}{2} \Omega_{im} x^m x^l - \tfrac{1}{4} \Omega_{il} |x|^2 \right) \partial_l \Psi \bigg) \Bigg]\ .
\end{align}
Then, $\partial_- \tensor{P}{^-_i} + \partial_j \tensor{P}{^j_i} = 0$ for each $i=1,2,3,4$.

\subsection{Type $VII$}

In six-dimensions the conformal Killing vector leads to the following diffeomorphism
\begin{align}
x^+&\to x^+ +\frac14\omega_2|x|^2\nonumber \\ 
x^-&\to x^- +\frac{1}{2} \omega_2 \left( x^- \right)^2 - \frac{1}{32}\omega_2 \omsq |x|^4 \nonumber \\
x^i &\to x^i +\frac{1}{8} \omega_2 \Omega_{ij} |x|^2 x^j + \frac{1}{2} \omega_2 x^i x^-
\ .
\end{align}
Upon reduction to five dimensions we find
\begin{align}
  \delta_{}  x^- &= \frac{1}{2} \omega_2 \left( x^- \right)^2 - \frac{1}{32}\omega_2 \omsq |x|^4\nonumber\\
  \delta_{}  x^i &= \frac{1}{8} \omega_2 \Omega_{ij} |x|^2 x^j + \frac{1}{2} \omega_2 x^i x^-\ .
  \end{align} 
Again the measure is rescaled
\begin{align}
	\delta_{} (dx^-d^4x) = 3\omega_2 x^- (dx^-d^4x)\ .
\end{align}
Following the discussion in appendix \ref{A1}   we find \begin{align}
  \delta_{}  A_- &= -\omega_2 x^- A_- - \frac{1}{2} \omega_2 x^i A_i \nonumber\\
  \delta_{}  A_i &=   -\frac{1}{2} \omega_2 x^- A_i   + \frac{1}{8} \omega_2 \omsq |x|^2 x^i A_- + \frac{1}{8} \omega_2 \left( \Omega_{ij} |x|^2 - 2\Omega_{jk} x^i x^k \right) A_j \nonumber\\
  \delta_{}  X^I &= -\omega_2 x^- X^I  \nonumber\\
  \delta_{}  G_{ij} &= - 2 \omega_2 x^- G_{ij} - \frac{1}{2}\left( \lambda^{ki} G_{kj} - \lambda^{kj} G_{ki} +\varepsilon_{ijkl} \lambda^{mk} G_{ml} \right) +\frac{1}{2} \omega_2  \left( x^i F_{-j} - x^j F_{-i} + \varepsilon_{ijkl} x^k F_{-l} \right)  \nonumber\\
  \delta_{}  \Psi &=  -\frac{1}{4}\omega_2 x^-(5+\Gamma_{-+})  \Psi+ \frac{1}{4} \omega_2 x^i \Gamma_{+}\Gamma_{i}\Psi + \frac{1}{4} \lambda^{ij}\Gamma_{ij}\Psi \ ,
  \end{align}
where 
\begin{align}
\lambda^{ij}=\frac{1}{4} \omega_2 \left( \Omega_{ik}x^k x^j - \Omega_{jk}x^k x^i \right) +\frac{1}{8}\omega_2\Omega_{ij}|x|^2 	\ .
\end{align} 
The conserved Noether current now takes the form
\begin{align} \label{km}
  K^- &= \left( \tfrac{1}{2} \left( x^- \right)^2 - \tfrac{1}{32}\omsq |x|^4 \right)\mathcal{L}_\Omega\nonumber\\
  & + \frac{1}{4\pi^2 \om}\text{tr}\,\Bigg[\,\, \tfrac{1}{32} |x|^2 \varepsilon_{ijkl} F_{ij} F_{kl} -\tfrac{1}{8} \omsq |x|^2 X^I X^I \nonumber\\
  &\hspace{12mm} + \left( F_{-i} + \tfrac{1}{2} \Omega_{jk} x^k G_{ij} \right) \bigg( -\tfrac{1}{2} x^- A_i + \tfrac{1}{8} \omsq |x|^2 x^i A_- \nonumber\\
  &\hspace{55mm} +\tfrac{1}{8} \left( \Omega_{il} |x|^2 - 2\Omega_{lm}x^m x^i \right) A_l \nonumber\\
  &\hspace{55mm} -\left( \tfrac{1}{2} \left( x^- \right)^2 - \tfrac{1}{32} \omsq |x|^4 \right)\partial_- A_i \nonumber\\
  &\hspace{55mm} -\left( \tfrac{1}{2} x^l x^- + \tfrac{1}{8} \Omega_{lm} x^m |x|^2 \right)\partial_l A_i \bigg) \nonumber\\
  &\hspace{12mm} -\tfrac{1}{2} \Omega_{ij} x^j \left( \nabla_i X^I \right) \bigg( x^- X^I + \left( \tfrac{1}{2} \left( x^- \right)^2 - \tfrac{1}{32} \omsq |x|^4 \right)\partial_- X^I \nonumber\\
  &\hspace{47mm} +\left( \tfrac{1}{2} x^k x^- + \tfrac{1}{8} \Omega_{kl} x^l |x|^2 \right)\partial_k X^I \bigg) \nonumber\\
  &\hspace{12mm}-\tfrac{\ii}{2}\bar{\Psi} \left( \Gamma_+ + \tfrac{1}{2} \Omega_{ij} x^j \Gamma_i \right)\bigg( -\tfrac{1}{4} x^- \left( 5+\Gamma_{-+} \right) \Psi + \tfrac{1}{4} x^k \Gamma_+ \Gamma_k \Psi \nonumber\\
  &\hspace{55mm} +\tfrac{1}{32}\left( 2\Omega_{km} x^m x^l - 2\Omega_{lm} x^m x^k + \Omega_{kl} |x|^2 \right)\Gamma_{kl}\Psi \nonumber\\
  &\hspace{55mm} -\left( \tfrac{1}{2} \left( x^- \right)^2 - \tfrac{1}{32} \omsq |x|^4 \right)\partial_- \Psi \nonumber\\
  &\hspace{55mm} -\left( \tfrac{1}{2} x^k x^- + \tfrac{1}{8} \Omega_{kl} x^l |x|^2 \right)\partial_k \Psi  \bigg) \Bigg] \ ,
\end{align}
\begin{align} \label{ki}
  K^i &= \left( \tfrac{1}{2} x^i x^- + \tfrac{1}{8} \Omega_{ij} |x|^2 x^j \right)\mathcal{L}_\Omega\nonumber\\
  & + \frac{1}{4\pi^2 \om}\text{tr}\,\Bigg[\,\, -\tfrac{1}{8} |x|^2 \varepsilon_{ijkl} F_{-j} F_{kl} +\tfrac{1}{4} \Omega_{ij} x^j X^I X^I \nonumber\\
  &\hspace{12mm} + \left( F_{-i} + \tfrac{1}{2} \Omega_{jk} x^k G_{ij} \right) \bigg( x^- A_- + \tfrac{1}{2} x^l A_l + \left( \tfrac{1}{2} \left( x^- \right)^2 - \tfrac{1}{32} \omsq |x|^4 \right)\partial_- A_-\nonumber\\
  &\hspace{55mm} + \left( \tfrac{1}{2} x^l x^- + \tfrac{1}{8} \Omega_{lm} x^m |x|^2 \right)\partial_l A_-\bigg)\nonumber\\
  &\hspace{12mm} + G_{ij} \bigg( -\tfrac{1}{2} x^- A_j + \tfrac{1}{8} \omsq |x|^2 x^j A_- +\tfrac{1}{8} \left( \Omega_{jk} |x|^2 - 2\Omega_{kl}x^l x^j \right) A_k  \nonumber\\
  &\hspace{27mm} -\left( \tfrac{1}{2} \left( x^- \right)^2 - \tfrac{1}{32} \omsq |x|^4 \right)\partial_- A_j -\left( \tfrac{1}{2} x^k x^- + \tfrac{1}{8} \Omega_{kl} x^l |x|^2 \right)\partial_k A_j \bigg) \nonumber\\
  &\hspace{12mm} + \left( \nabla_i X^I \right) \bigg( x^- X^I + \left( \tfrac{1}{2} \left( x^- \right)^2 - \tfrac{1}{32} \omsq |x|^4 \right)\partial_- X^I \nonumber\\
  &\hspace{35mm} +\left( \tfrac{1}{2} x^j x^- + \tfrac{1}{8} \Omega_{jk} x^k |x|^2 \right)\partial_j X^I \bigg) \nonumber\\
  &\hspace{12mm}+\tfrac{\ii}{2}\bar{\Psi} \Gamma_i\bigg( -\tfrac{1}{4} x^- \left( 5+\Gamma_{-+} \right) \Psi + \tfrac{1}{4} x^k \Gamma_+ \Gamma_k \Psi \nonumber\\
  &\hspace{30mm} +\tfrac{1}{32}\left( 2\Omega_{km} x^m x^l - 2\Omega_{lm} x^m x^k + \Omega_{kl} |x|^2 \right)\Gamma_{kl}\Psi \nonumber\\
  &\hspace{30mm} -\left( \tfrac{1}{2} \left( x^- \right)^2 - \tfrac{1}{32} \omsq |x|^4 \right)\partial_- \Psi \nonumber\\
  &\hspace{30mm} -\left( \tfrac{1}{2} x^k x^- + \tfrac{1}{8} \Omega_{kl} x^l |x|^2 \right)\partial_k \Psi  \bigg) \Bigg]\ .
\end{align}

\section{Conclusion}\label{Conclusion}

In addition to enjoying 16 supersymmetries and 16 superconformal symmetries, the $(2,0)$ theory is invariant under 6 translations and the 15 generators of the six-dimensional Lorentz group, {\it i.e.} the 21 generators of the six-dimensional Poincar\'{e} group. In addition there are 6 special conformal symmetries and 1 dilatation symmetry. The bosonic symmetries are then just those of the six-dimensional conformal group $SO(2,6)$. In total these comprise 32 fermionic and 28 bosonic symmetries. We would like a description of the $(2,0)$ theory that has as many of these symmetries manifest as possible. 

Although the $(2,0)$ theory does not appear to have a  six-dimensional lagrangian description in general, much can be learned by reducing the theory to five dimensions, whereupon we obtain five-dimensional super-Yang-Mills theory which has been conjectured to provide a complete description of the $(2,0)$ theory nonperturbatively \cite{Douglas:2010iu,Lambert:2010iw}. By reducing on a spacelike (or timelike) circle we break all conformal  and superconformal symmetries and reduce the six-dimensional Poincar\'{e} group to the five-dimensional one with 15 generators. In addition we still have translations in the compact direction as a symmetry too (albeit trivially but one can still identify a conserved charge as the topological instanton number). Thus we find 16 supersymmetries and 16 bosonic spacetime symmetries.

If we instead reduce on a null direction, then we preserve 16 supersymmetries, 8 super-conformal symmetries, the 10 symmetries of the four-dimensional euclidean group, as well as translations in $x^-$, a scale transformation and the trivial translation in the reduced null direction. We find that there are also 4 null boosts (type $VI$) and an additional conformal symmetry (type $VII$). Thus we find 24 (conformal) supersymmetries and 18 bosonic symmetries. On the other hand, if we place the $(2,0)$ theory on the spacetime (\ref{gdef}) which was deduced by writing $AdS_7$ as a timelike fibration over a non-compact complex projective space $\tilde{\CP}^3$ and going to the boundary, then reducing along the null direction will give an $\Omega$-deformed theory with 8 supersymmetries and 16 superconformal symmetries. We also find the 15 bosonic symmetries of $SU(3,1)$ (type $II$ to $VII$) expected from holography, and the trivial translation along the reduced null direction (type $I$). Thus we find 24 (conformal) supersymmetries and 16 bosonic symmetries. Although we have lost two rotational symmetries compared with straightforward null reduction there is an additional benefit that we maintain a more direct link to the non-compact theory and $AdS$ dual.

The existence of five-dimensional lagrangians with such high degrees of symmetry is clearly remarkable. We therefore plan to investigate the following questions in order to elucidate their mathematical structure and physical significance: 

\begin{itemize} 

\item In addition to having 24 supercharges and an $SU(3,1)$ symmetry, the theories we consider have an $Sp(4) \sim SO(5)$ R-symmetry corresponding to the isometries of the $S^4$ in the bulk geometry (this symmetry becomes manifest if we write our 32-component spinors as 8-component spinors with $Sp(4)$ indices). It is therefore natural to combine all of these symmetries into a supergroup whose bosonic subgroup is $SU(3,1) \times Sp(4)$. This is not a superconformal group since $SU(3,1)$ is not equivalent to $SO(p,q)$ for any $p+q=8$, but it seems to be a Wick rotation of the supergroup $OSp(6|4)$, which is enjoyed by the ABJM theory and admits an infinite-dimensional extension known as Yangian symmetry (the superconformal group of $\mathcal{N}=4$ super-Yang-Mills also exhibits such an extension, see \cite{Beisert:2010jr} for a review). It would therefore be interesting to investigate the supergroup structure of null reductions of the $(2,0)$ theory and the possibility of an infinite-dimensional extension. A proposal for seeing Yangian symmetry at the lagrangian level was recently described in \cite{Beisert:2018zxs} and demonstrated for $\mathcal{N}=4$ super-Yang-Mills and the ABJM theory. 

\item Demonstrating $SU(3,1)$ symmetry of the $\Omega$-deformed null reduction of the $(2,0)$ theory provides an important test of the holographic duality, but it would be desirable to go beyond matching symmetries by probing dynamics. As shown in \cite{Lambert:2011gb ,Lambert:2018lgt,Mouland:2019zjr}, the dynamics of the $\Omega_{ij}=0$ theories can be reduced to quantum mechanics on the moduli space of instantons. Noting that the instantons correspond to Kaluza-Klein modes along the null direction, it would therefore be interesting to work out the quantum mechanical description of the $\Omega$-deformed null reduction in the limit that the rank of the gauge group goes to infinity and match it with the action for D0-branes in $\tilde{\CP}^3 \times S^4$. Another important test of the duality would be to compute correlation functions and match them with Witten diagrams in the bulk. Two-point functions of chiral primary operators were computed in the original DLCQ proposal \cite{Aharony:1997an}, although the extension to higher-point functions appears to be challenging. On the other hand, having a field theory description should make such calculations more tractable.

\item The metric in \eqref{gdef} is a conformal compactification of six-dimensional Minkowski space which can be generalised to other dimensions. It would therefore be of interest to perform a similar null reduction of other superconformal field theories, such as four-dimensional  $\mathcal{N}=2$ super-Yang-Mills coupled to suitable matter and  the $\mathcal{N}=4$ theory. In this case, one would put the theory on the following conformal compactification of Minkowski space:
\begin{align}
	ds^{2}\ =\ \frac{-2dx^{+}\left(dx^{-}-\frac{\ii}{2\om}(zd\bar z - \bar z d z)\right)+dzd\bar z}{\cos^{2}\left(x^{+}/2\om\right)}\ ,
\end{align}
where 
$z =x^1+\ii x^2$ and 
$\Omega_{12}=-\Omega_{21}=1/\om$. Again for a conformal field theory we can neglect the denominator. The corresponding reduction in  IIB string theory  would involve constructing $AdS_5$ as timelike fibration over $\tilde{\CP}^2$. Although not necessary we could again reduce along the null direction to find an $\Omega$-deformed three-dimensional Yang-Mills theory. Such a reduction would break all the supersymmetry \cite{Pope:1999xg}, unless a suitable twisting by the R-symmetry can be introduced. It would nevertheless be interesting to see how the well-known holographic dictionary becomes modified, and how various important properties of super-Yang-Mills such as integrability and S-duality are encoded in the three-dimensional description. It may also be possible to relate this to the chiral algebra description of four-dimensional superconformal field theory theories proposed in \cite{Beem:2013sza}. This would provide new insight into null reductions of the $(2,0)$  theory and other conformal field theories and would also be interesting in its own right. 

\end{itemize}

Ultimately, we hope that pursuing these directions will further our understanding of the underlying dynamics of M-theory.

\section*{Acknowledgement}

N.~Lambert and P.~Richmond were supported   by STFC grant ST/L000326/1, A.~Lipstein by the Royal Society as a Royal Society University Research Fellowship holder and  R.~Mouland by the STFC studentship ST10837. 

\appendix

\section{A Six-Dimensional Origin for the Symmetries }\label{A1}

In this appendix we  provide a six-dimensional origin for the symmetries found above. Of course the main problem is that there is no known lagrangian for the $(2,0)$ theory in six dimensions, nor is there expected to be one. However    let us consider the following action
\begin{align}
  S_{6D} = \frac{1}{8\pi^3 \om^2 } \text{tr} \int d^6 x\,\, \sqrt{-g} \bigg\{ &-\frac{1}{12} H_{\mu\nu\rho}H^{\mu\nu\rho} - \frac{1}{2} g^{\mu\nu} D_\mu X^I D_\nu X^I \nonumber\\
  & +\frac{\ii}{2} \bar{\Psi} \Gamma^\mu {\cal D}_\mu \Psi -\frac{1}{2} V^\mu \bar{\Psi} \Gamma_\mu \Gamma^I \left[ X^I, \Psi \right]  \bigg\}\ .
\end{align}
Note that in this appendix we use $\Gamma_\mu$ to denote six-dimensional curved space $\Gamma$-matrices. In the other sections of the paper all $\Gamma$-matrices are those of Minkowski space and as such can be identified with the tangent frame $\Gamma$-matrices that appear  in this appendix. Furthermore
here $\mu=\{+,-,i\}$ and we have introduced a three-form $H_{\mu\nu\rho}$ and vector field $V^\mu$. 

We emphasise that we are not proposing $S_{6D}$ as a candidate for the $(2,0)$ theory. Rather we merely wish to use it to motivate the symmetries of the reduced theory we discussed in the main text above. In particular we will use two features of $S_{6D}$: it has six-dimensional diffeomorphism invariance and, using a suitable ansatz,  it can be dimensionally reduced to $S_\Omega$, up to a single topological term whose variation is a total derivative. We will then see that the somewhat unusual transformations we used above have a more standard interpretation  within the context of $S_{6D}$.

We have a vielbein $e^{\underline{\mu}}{}_\mu$ satisfying $e^{\underline{\mu}}{}_\mu \eta_{\underline{\mu}\underline{\nu}}e^{\underline{\nu}}{}_\nu=g_{\mu\nu}$, where we choose lightcone coordinates in the tangent frame, {\it i.e.}\ $\eta_{\underline{+}\underline{-}}=-1, \eta_{\underline{+}\underline{+}}=\eta_{\underline{-}\underline{-}}=0, \eta_{\underline{i}\underline{j}}=\delta_{ij}$. Then, we have $\Gamma^\mu = e^{\mu}{}_{\underline{\mu}}\Gamma^{\underline{\mu}}$ and $\Gamma^I = \delta^I_{\und{I}} \Gamma^{\und{I}}$, where $\left\{ \Gamma^{\underline{\mu}}, \Gamma^{\underline{I}} \right\}$ form a (real) basis for the eleven-dimensional Clifford algebra.  The gauge covariant derivative   is $ D_\mu  = \partial_\mu  - \ii \left[ A_\mu, \,. \,\, \right]$, while on $\Psi$  we have
\begin{align}
  {\cal D}_\mu \Psi = \left( D_\mu + \frac{1}{4}  \omega_\mu^{\underline{\mu\nu} }\Gamma_{\underline{\mu\nu}} \right) \Psi \ .
\end{align}

By construction $S_{6D}$ is invariant under six-dimensional diffeomorphisms. In particular
given a vector field ${k}^\mu$, the infinitesimal diffeomorphism generated by $k^\mu$ is given by
\begin{align}
  \delta_d x^\mu &= {k}^\mu \nonumber\\
  \delta_d \tensor{T}{^{\mu_1}^{\dots}^{\mu_r}_{\nu_1}_{\dots}_{\nu_s}} &=   \left( \partial_\rho {k}^{\mu_1} \right) \tensor{T}{^{\rho}^{\mu_2}^{\dots}^{\mu_r}_{\nu_1}_{\dots}_{\nu_s}} + \dots \nonumber\\
  &\hspace{12mm} - \left( \partial_{\nu_1} {k}^{\rho}  \right) \tensor{T}{^{\mu_1}^{\dots}^{\mu_r}_{\rho}_{\nu_2}_{\dots}_{\nu_s}} - \dots \nonumber\\
  &= - \tensor{\left(\mathcal{L}_{{k}} T \right)}{^{\mu_1}^{\dots}^{\mu_r}_{\nu_1}_{\dots}_{\nu_s}} + {k}^\rho \partial_\rho \tensor{T}{^{\mu_1}^{\dots}^{\mu_r}_{\nu_1}_{\dots}_{\nu_s}} \nonumber\\
  \delta_d \Psi &= \frac{1}{4} {\lambda}^{\und{\mu\nu}} \Gamma_{\und{\mu\nu}} \Psi \nonumber\\
  \delta_d e^{\und{\mu}}_{\mu} &= - \left( \partial_\mu {k}^\rho \right) e^{\und{\mu}}_{\rho} +  \tensor{{\lambda}}{^{\und{\mu}}_{\und{\nu}}} e^{\und{\nu}}_{\mu} \nonumber\\
  \delta_d \tensor{\omega}{_\mu^{\und{\mu\nu}}} &= -\left( \partial_\mu {k}^\rho \right) \tensor{\omega}{_\rho^{\und{\mu\nu}}} + \tensor{{\lambda}}{^{\und{\mu}}_{\und{\rho}}}\, \tensor{\omega}{_\mu^{\und{\rho\nu}}} + \tensor{{\lambda}}{^{\und{\nu}}_{\und{\rho}}}\, \tensor{\omega}{_\mu^{\und{\mu\rho}}} - \partial_\mu {\lambda}^{\und{\mu\nu}}\ ,
\end{align}
where ${T}^{\mu_1 \dots \mu_r}_{\nu_1 \dots \nu_s}$ is a general $(r,s)$-tensor, and we've allowed for a local infinitesimal Lorentz transformation $\tensor{{\lambda}}{^{\und{\mu}}_{\und{\nu}}}$ in the tangent frame. We are   assuming  here that the components of ${k}^\mu$ in a given coordinate frame are small so that we need only consider the first order terms. Note also that we are here regarding the diffeomorphism as a \textit{passive} transformation.

Next we want to write  $S_{6D}$ explicitly in a coordinate frame in which the metric is given by (\ref{gdef}). This metric admits the choice of vielbein $e^{\und{+}}{}_{+} = 1, e^{\und{-}}{}_{-} = 1,  e^{\und{-}}{}_{i}=\frac{1}{2} \Omega_{ij}x^j$ and $e^{\und{i}}{}_{j} = \delta_{ij}$, with all other components vanishing. We suppose that the vector $V^\mu$ takes the form $V^+=1$ with all other components vanishing. Furthermore we choose to turn off any $x^+$ dependance of the fields, and set $A_+=0$, in turn implying $F_{+\mu}=0$. We can then make the identification
\begin{align}\label{FH}
F_{\mu\nu} = H_{\mu\nu +}	\ .
\end{align}
To match with the actions above we  define
\begin{align}\label{GH}
G_{ij}=H_{-ij}+\frac{1}{2} \varepsilon_{ijkl}H_{-kl} \ .
\end{align}
After performing the trivial $x^+$ integral, we find that $S_\Omega$  agrees with the reduced $S_{6D}$   up to two additional terms:
\begin{align}\label{SS}
  S_{\Omega} &= \frac{1}{4\pi^2 \om}\text{tr} \int d^5x\, \bigg\{ \frac12 F_{-i}F_{-i} - \frac12  {\nabla}_i X^I {\nabla}_i X^I + \frac12 {\cal F}_{ij}G_{ij} \nonumber\\
  &\hspace{33mm}  -\frac{\ii}{2}\bar\Psi\Gamma_{\und{+}}D_-\Psi + \frac{\ii}{2}\bar\Psi\Gamma_{\und{i}} {\nabla}_i\Psi - \frac{1}{2}\bar\Psi\Gamma_{\und{+}}\Gamma^{I}[X^I,\Psi] \bigg\}\nonumber\\
  &= S_{6D} + \frac{1}{4\pi^2 \om} \text{tr} \int d^5x\, \bigg\{ \frac{1}{4} \varepsilon_{ijkl} \mathcal{F}_{ij} H_{-kl} \nonumber\\
  &\hspace{45mm}+ \frac{1}{12} \left( H_{ijk} + \frac{3}{2} \Omega_{l[i|}H_{-|jk]} x^l \right) \Big( H_{ijk} + \frac{3}{2} \Omega_{m[i|}H_{-|jk]} x^m \Big)  \bigg\} \ ,\end{align}
where, as above,  $ {\nabla}_i = D_i -\frac12\Omega_{ij}x^jD_-$ and ${\cal F}_{ij}  = F_{ij} - \frac12\Omega_{ik}x^kF_{-j}+ \frac12 \Omega_{jk}x^kF_{-i}$. 
Lastly we can impose the relation
\begin{align}\label{Hijk}
H_{ijk}= - \frac{3}{2} \Omega_{l[i|}H_{-|jk]} x^l\ . 
\end{align}
This ensures that the second line in  (\ref{SS}) vanishes and as such we   have  
\begin{align}
  S_{\Omega}  =S_{6D} +  \frac{1}{4\pi^2\om}   \text{tr}  \int d^5x\, & \frac{1}{4} \varepsilon_{ijkl} \mathcal{F}_{ij} H_{-kl} \ .
  \end{align}
Note that (\ref{Hijk}) differs from that used in the construction of \cite{Lambert:2019jwi}. However we emphasise again that $S_{6D}$ should not be taken literally as an action for the $(2,0)$ theory. In particular with the ansatz here $H_{\mu\nu\lambda}$ is not self-dual.

We now wish to construct a bosonic symmetry $\delta$ for $S_\Omega$ that descends from the diffeomorphisms for $S_{6D}$. In particular we start with a natural guess $\delta_\text{trial}$ that comes from diffeomorphisms which we then need to slightly correct using the scaling symmetry to find the total variation $\delta_{}$. For a generic object $\Phi$, we are free to replace $\Phi$ in $S_{6D}$ with an explicit expression $\Phi(x)$ in some coordinate frame and preserve a passive diffeomorphism ${k}^\mu$ only if we have
\begin{align}
  \hat{\delta} \Phi := {k}^\rho \partial_\rho \Phi - \delta_d \Phi = 0\ .
  \label{eq: tilde hat def}
\end{align}
In other words the transformation of $\Phi$, as induced by its dependence on $x^\mu$, must match its transformation under $\delta_d$. For a tensor field $T$, we have $\hat{\delta} T = \mathcal{L}_{{k}} T$, and so for ${k}^\mu$ Killing, we have $\hat{\delta}g_{\mu\nu} = 0$. We will consider instead the more general space of conformal Killing vectors with $\partial_+ {k}^\mu=0$, contained within $k^\mu$ as given in (10). These satisfy $\mathcal{L}_k g_{\mu\nu} = \omega g_{\mu\nu}$, with $\omega= \omega_1 + \Omega_{ij} v_i x^j + \omega_2 x^-$.
So we choose to replace $\{g_{\mu\nu},e^{\und{\mu}}{}_{\mu},\tensor{\omega}{_\mu^{\und{\mu \nu}}}, V^\mu \}$ with their coordinate expressions. Then, $\delta_\text{trial}$ is defined to act as $k^\rho\partial_\rho$ on these fields, and as $\delta_d$ on everything else. Equivalently, we have $\delta_\text{trial}=\delta_d + \hat{\delta}$, where $\hat{\delta}$ as defined in (\ref{eq: tilde hat def}) acts only on $\{g_{\mu\nu},e^{\und{\mu}}{}_{\mu},\tensor{\omega}{_\mu^{\und{\mu  \nu}}}, V^\mu \}$.

As we've already seen, we have $\hat{\delta}g_{\mu\nu}=\omega g_{\mu\nu}$. Next, we note that the conformal Killing equation implies that
\begin{align}
  k^\rho \partial_\rho e^{\und{\mu}}{}_{\mu} + \left( \partial_\mu k^\rho \right) e^{\und{\mu}}{}_{\rho} = \tensor{\lambda}{^{\und{\mu}}_{\und{\nu}}} e^{\und{\nu}}{}_{\mu} + \tfrac{1}{2} \omega e^{\und{\mu}}{}_{\mu}\ ,
\end{align}
for local Lorentz transformation $\tensor{\lambda}{^{\und{\mu}}_{\und{\nu}}}$ given by
\begin{align}
  \tensor{\lambda}{^{\und{\mu}}_{\und{\nu}}} = \left( \partial_\mu k^\nu \right) e_{\und{\nu}}{}^{\mu} e^{\und{\mu}}{}_{\nu} + k^\nu e_{\und{\nu}}{}^{\mu} \partial_\nu e^{\und{\mu}}{}_{\mu} - \tfrac{1}{2} \omega \delta^{\und{\mu}}_{\und{\nu}}\ .
\end{align}
One can show using the conformal Killing equation that this does indeed satisfy $\tensor{\lambda}{_{\und{\mu \nu}}}+\tensor{\lambda}{_{\und{\nu \mu}}}=0$. Then, choosing this $\tensor{\lambda}{_{\und{\mu \nu}}}$ for the diffeomorphism $\delta_d$, we have $\hat{\delta}e^{\und{\mu}}{}_{\mu}=\tfrac{1}{2} \omega e^{\und{\mu}}{}_{\mu}$. Next we find that for the spin connection term we have
\begin{align}
  \hat{\delta}\left( \frac{1}{4} \bar{\Psi} \Gamma^{{\mu}}\, \tensor{\omega}{_\mu^{\und{\nu\rho}}} \Gamma_{\und{\nu\rho}}\Psi \right) = 0\ .
\end{align}
Finally, we simply have $\hat{\delta}V^\mu = 0$.

To continue we observe that
\begin{align}\label{dSS}
  \delta_\text{trial} S_\Omega  = \hat{\delta} S_{6D}
    + \frac{1}{4\pi^2\om} \delta_\text{trial}\Bigg[ \text{tr}  \int d^5x\, & \frac{1}{4} \varepsilon_{ijkl} \mathcal{F}_{ij} H_{-kl}   \Bigg]\ , \nonumber
\end{align}
where we have used $\delta_d S_{6D}=0$.   Note that once we impose (\ref{Hijk}) it is not necessary to also require that 
\begin{align}
	\delta_\text{trial}\left[H_{ijk}+ \frac{3}{2} \Omega_{l[i|}H_{-|jk]} x^l\right]=0\ ,
\end{align} 
to ensure that the variation of the  second line in (\ref{SS}) vanishes since the right hand side is quadratic in $H_{ijk}+ \tfrac{3}{2} \Omega_{l[i|}H_{-|jk]} x^l$. We also do not need to worry about the relation (\ref{GH}) as this defines $G_{ij}$ and hence will define its variation.

However  we do require that the identification (\ref{FH}) is consistent with the diffeomorphism. Under a general diffeomorphism ${k}^\mu$   we have
\begin{align}
  \delta_\text{trial} H_{+\mu\nu} &= -( \partial_\mu k^\lambda ) H_{+\lambda\nu} - ( \partial_\nu k^\lambda ) H_{+\mu\lambda} -(\partial_+ {k}^\lambda )H_{\lambda\mu\nu}\nonumber\\
  \delta_\text{trial} F_{\mu\nu} &= -( \partial_\mu k^\lambda ) F_{\lambda\nu} - ( \partial_\nu k^\lambda ) F_{\mu\lambda}\ .
\end{align}
We see that $\delta_\text{trial} F_{\mu\nu} = \delta_\text{trial} H_{+\mu\nu}$ only if $\partial_+ {k}^\mu=0$ and so (\ref{FH})  is invariant under this restricted set of diffeomorphisms. Unsurprisingly this breaks the space of symmetries to those ${k}^\mu$ and $\tensor{{\lambda}}{^{\und{\mu}}_{\und{\nu}}}$ that are independent of $x^+$.

Thus we are led to the $\mathcal{F}_{ij}H_{-kl}$ term. We find
  \begin{align}
  \delta_\text{trial} \mathcal{F}_{ij} &= - \omega \mathcal{F}_{ij} - \left( \delta_i^{\und{i}} \tensor{\lambda}{^{\und{k}}_{\und{i}}}\delta^k_{\und{k}} \right) \mathcal{F}_{kj} + \left( \delta_j^{\und{j}} \tensor{\lambda}{^{\und{k}}_{\und{j}}}\delta^k_{\und{k}} \right) \mathcal{F}_{ki} \nonumber\\
  \delta_\text{trial} H_{-kl} &= -2\omega H_{-kl} - \left( \delta_k^{\und{k}} \tensor{\lambda}{^{\und{i}}_{\und{k}}}\delta^i_{\und{i}} \right) H_{-il} + \left( \delta_l^{\und{l}} \tensor{\lambda}{^{\und{i}}_{\und{l}}}\delta^i_{\und{i}} \right) H_{-ik} \nonumber \\
  & \qquad +\frac{1}{2}\left( 2 v_k + \omega_2 x^k \right)F_{-l} - \frac{1}{2}\left( 2 v_l + \omega_2 x^l \right)F_{-k}\ .
\end{align}
Indeed, these forms follow almost immediately when one notes the forms of $\mathcal{F}_{ij}$ and $H_{-kl}$ in terms of tangent frame fields; $\mathcal{F}_{ij} = H_{\und{+ij}}$, $H_{-kl}=H_{\und{-kl}}$. Then, noting that $\delta_\text{trial}\left( d^5 x \right)=3\omega d^5 x$ and that the local Lorentz pieces exactly vanish, we find
\begin{align}
 \frac{1}{4\pi^2\om}  \delta_\text{trial}\Bigg[ \text{tr} \int d^5x\, \bigg( \frac{1}{4} \varepsilon_{ijkl} \mathcal{F}_{ij} H_{-kl} \bigg)  \Bigg] =  \frac{1}{4\pi^2\om} \text{tr} \int d^5x\, \frac{1}{2}\varepsilon_{ijkl} \left( v_k + \frac{1}{2} \omega_2 x^k  \right) F_{ij} F_{-l}\ .
\end{align}
This term is essentially $d k^+\wedge \text{tr} \left( F\wedge F \right)$, and so is a total derivative. In particular, we have
\begin{align}
  \varepsilon_{ijkl} \left( v_k + \frac{1}{2} \omega_2 x^k  \right) \text{tr} \left( F_{ij} F_{-l} \right) =& - \partial_- \Big(  k^+  \varepsilon_{ijkl} \text{tr} \left( F_{ij} F_{kl} \right) \Big) \nonumber\\
  &+4\partial_i \Big(  k^+   \varepsilon_{ijkl} \text{tr} \left( F_{-j} F_{kl} \right) \Big)\ .
\end{align}
Hence we are left with
\begin{align}
  \delta_\text{trial} S_\Omega = \hat{\delta} S_{6D} = \frac{1}{4\pi^2\om} \text{tr} \int d^5 x\, \bigg\{ &2\omega \left( -\frac{1}{2} {\nabla}_i X^I {\nabla}_i X^I \right) \nonumber \\
  &+ \frac{5}{2}\omega \left( -\frac{\ii}{2} \bar{\Psi} \Gamma_{\und{+}}D_- \Psi + \frac{\ii}{2} \bar{\Psi} \Gamma_{\und{i}}  {\nabla}_i \Psi \right) \nonumber\\
  &+ \frac{7}{2} \omega \left( -\frac{1}{2} \bar{\Psi} \Gamma_{\und{+}} \Gamma^{I} \left[ X^I, \Psi\right] \right) \bigg\}\ .
  \label{eq: S variation before scaling}
\end{align}
Lastly if we  augment $\delta_\text{trial}$ by a simple scaling by $\omega$
\begin{align}
  \delta' X^I &= - \omega X^I \nonumber\\
  \delta' \Psi &= -\frac{5}{4} \omega \Psi\ ,
 \end{align}
then for $\delta_{}= \delta_\text{trial} + \delta'$, we have $\delta_{}S_\Omega=0$.

In summary  we have
\begin{align}
  \delta_{} x^\mu &= k^\mu \nonumber\\
  \delta_{} A_- &= -\left( \partial_- k^- \right) A_- - \left( \partial_- k^i \right) A_i \nonumber\\
  \delta_{} A_i &= -\left( \partial_i k^- \right) A_- - \left( \partial_i k^j \right) A_j \nonumber\\
  \delta_{} X^I &= -\omega X^I \nonumber\\
 \delta_{} G_{ij} &= -2\omega G_{ij} - \frac{1}{2} \left( \delta_i^{\und{i}} \tensor{\lambda}{^{\und{k}}_{\und{i}}}\delta^k_{\und{k}} \right) G_{kj} + \frac{1}{2} \left( \delta_j^{\und{j}} \tensor{\lambda}{^{\und{k}}_{\und{j}}}\delta^k_{\und{k}} \right) G_{ki} - \frac{1}{2} \varepsilon_{ijkl} \left( \delta_k^{\und{k}} \tensor{\lambda}{^{\und{m}}_{\und{k}}}\delta^m_{\und{m}} \right) G_{ml} \nonumber \\
  &\quad + \frac{1}{2} \left( 2v_i + \omega_2 x^i \right) F_{-j} - \frac{1}{2} \left( 2v_j + \omega_2 x^j \right) F_{-i} + \frac{1}{2} \varepsilon_{ijkl}\left( 2v_k + \omega_2 x^k \right) F_{-l}\nonumber\\ 
  \delta_{} \Psi &= -\frac{5}{4} \omega \Psi +  \frac{1}{4} \lambda^{\und{\mu \nu}} \Gamma_{\und{\mu \nu}} \Psi \ ,
\end{align}
where $\lambda^{\und{\mu \nu}}=-\lambda^{\und{\nu \mu}}$ and
\begin{align}
  \lambda^{\und{-+}} &= -\frac{1}{2}\omega \nonumber\\
  \lambda^{\und{-i}} &= 0 \nonumber\\
  \lambda^{\und{+i}} &= v_i + \frac{1}{2} \omega_2 x^i \nonumber\\
  \lambda^{\und{ij}} &= M_{ij} + \frac{1}{2}\left( \Omega_{ij} v_k x^k + \Omega_{ik} v_k x^j - \Omega_{jk} v_k x^i + \Omega_{ik} v_j x^k - \Omega_{jk} v_i x^k \right) \nonumber\\
  &\qquad + \frac{1}{8} \omega_2 \left( \Omega_{ij} |x|^2 + 2 \left( \Omega_{ik} x^k x^j - \Omega_{jk} x^k x^i \right)\right)\ . 
\end{align}
Given the  form for $k^\mu$ specified in (\ref{kis}) we can compute an explicit expression for $\delta_{}$. 
From the point view of the five-dimensional field theory one can decompose  $\delta_{}$ into a diffeomorphism contribution, a scale transformation (of the form (\ref{weights})) as well as a tensor-like transformation that mixes the various components of the fields.

\section{$SU(3,1)$ Symmetry} \label{A2}

In this appendix, we will derive the conformal Killing vectors of the boundary metric from bulk Killing vectors following the method in \cite{Hoxha:2000jf,Huang:2010qy}, which makes the underlying $SU(3,1)$ symmetry manifest. The embedding coordinates for $AdS_{7}$ with unit radius satisfy
\begin{equation}
\bar{Z}\cdot \tilde \eta\cdot Z=-1\label{embed}\ ,
\end{equation}
where $Z^{I}\in\left\{ Z^{0},...,Z^{3}\right\} $ and $\tilde \eta={\rm diag}(-1,1,1,1)$.
The embedding coordinates can be written in terms of $x^{\mu}=\left(x^{+},x^{-},x^{i},\phi\right)$
as follows \cite{Pope:1999xg}\footnote{Our coordinates are related to the ones in \cite{Pope:1999xg} as follows: $x^+ =  \tau$, $x_3= y_1$, $x_4=y_2$, $ x^-=\chi - \frac{1}{2}(x_1 y_1+x_2 y_2)$.}:
\begin{align}
Z^{0}&=e^{\ii x^{+}/2}\left(\cosh\phi/2+\frac{1}{2}e^{\phi/2}\left(\ii x^{-}+\frac{1}{4}x_{i}^{2}\right)\right)
\nonumber\\
Z^{1}&=\frac{1}{2}e^{\left(\phi+\ii x^{+}\right)/2}\left(x_{1}+\ii x_{3}\right)\nonumber\\
Z^{2}&=\frac{1}{2}e^{\left(\phi+\ii x^{+}\right)/2}\left(x_{2}+\ii x_{4}\right)
\nonumber\\
Z^{3}&=e^{\ii x^{+}/2}\left(\sinh\phi/2-\frac{1}{2}e^{\phi/2}\left(\ii x^{-}+\frac{1}{4}x_{i}^{2}\right)\right)
\ ,
\end{align}
and the metric is given by

\begin{equation}
ds^{2}=d\bar{Z}\cdot\tilde \eta\cdot dZ\ .\label{adsmetric}
\end{equation}
Moreover the Kahler form is given by $J=dA$ where
\begin{align}
A=-\ii\bar{Z}\cdot\tilde \eta\cdot dZ\ .
\end{align}
After reducing along the timelike fibre parameterized by $x^+$, the $SO(6,2)$ symmetry is
broken to $SU(3,1)$, which is manifest in \eqref{embed}. The Killing vectors
associated with the remaining symmetries can be determined from \begin{align}
K_{A}^{\mu}=J^{\mu\nu}\partial_{\nu}\omega_{A}\ ,
\end{align}
where indices of the Kahler form are raised using the metric in \eqref{adsmetric}. Here $\omega_{A}$ are  the 15 scalar
functions
\begin{align}
\omega_{A}=\bar{Z}\cdot\tilde \eta\cdot T_{A}\cdot Z,\qquad A=1,...,15\ ,
\end{align} 
 and  $T_{A}$ are the generators of $SU(3,1)$:
\begin{align}
T_{1}&=\left(\begin{array}{cccc}
0 & 1 & 0 & 0\\
1 & 0 & 0 & 0\\
0 & 0 & 0 & 0\\
0 & 0 & 0 & 0
\end{array}\right)\ \ \ \ \ \ \ \  T_{2} =\left(\begin{array}{cccc}
0 & -\ii & 0 & 0\\
\ii & 0 & 0 & 0\\
0 & 0 & 0 & 0\\
0 & 0 & 0 & 0
\end{array}\right)\ \ \ \ \ \ \ T_{3}=\left(\begin{array}{cccc}
1 & 0 & 0 & 0\\
0 & -1 & 0 & 0\\
0 & 0 & 0 & 0\\
0 & 0 & 0 & 0
\end{array}\right) 
\nonumber\\
T_{4}&=\left(\begin{array}{cccc}
0 & 0 & 1 & 0\\
0 & 0 & 0 & 0\\
1 & 0 & 0 & 0\\
0 & 0 & 0 & 0
\end{array}\right) \ \ \ \ \ \ \ \ T_{5} =\left(\begin{array}{cccc}
0 & 0 & -\ii & 0\\
0 & 0 & 0 & 0\\
\ii & 0 & 0 & 0\\
0 & 0 & 0 & 0
\end{array}\right)\ \ \ \ \ \ \ T_{6}=\left(\begin{array}{cccc}
0 & 0 & 0 & 0\\
0 & 0 & 1 & 0\\
0 & 1 & 0 & 0\\
0 & 0 & 0 & 0
\end{array}\right) 
\nonumber\\
T_{7}&=\left(\begin{array}{cccc}
0 & 0 & 0 & 0\\
0 & 0 & -\ii & 0\\
0 & \ii & 0 & 0\\
0 & 0 & 0 & 0
\end{array}\right)\ \ \ \ \ \  T_{8} =\frac{1}{\sqrt{3}}\left(\begin{array}{cccc}
1 & 0 & 0 & 0\\
0 & 1 & 0 & 0\\
0 & 0 & -2 & 0\\
0 & 0 & 0 & 0
\end{array}\right)
\ \  T_{9}=\left(\begin{array}{cccc}
0 & 0 & 0 & 1\\
0 & 0 & 0 & 0\\
0 & 0 & 0 & 0\\
-1 & 0 & 0 & 0
\end{array}\right)
\nonumber\\
T_{10}&=\left(\begin{array}{cccc}
0 & 0 & 0 & -\ii\\
0 & 0 & 0 & 0\\
0 & 0 & 0 & 0\\
-\ii & 0 & 0 & 0
\end{array}\right)\ \ \   T_{11} =\left(\begin{array}{cccc}
0 & 0 & 0 & 0\\
0 & 0 & 0 & 1\\
0 & 0 & 0 & 0\\
0 & -1 & 0 & 0
\end{array}\right) \ \ \ \ \ \ T_{12}=\left(\begin{array}{cccc}
0 & 0 & 0 & 0\\
0 & 0 & 0 & -\ii\\
0 & 0 & 0 & 0\\
0 & -\ii & 0 & 0
\end{array}\right)
\nonumber\\
T_{13}&=\left(\begin{array}{cccc}
0 & 0 & 0 & 0\\
0 & 0 & 0 & 0\\
0 & 0 & 0 & 1\\
0 & 0 & -1 & 0
\end{array}\right)\ \ \ \  T_{14}=\left(\begin{array}{cccc}
0 & 0 & 0 & 0\\
0 & 0 & 0 & 0\\
0 & 0 & 0 & -\ii\\
0 & 0 & -\ii & 0
\end{array}\right)\ \ \ \  T_{15}=\frac{1}{\sqrt{6}}\left(\begin{array}{cccc}
1 & 0 & 0 & 0\\
0 & 1 & 0 & 0\\
0 & 0 & 1 & 0\\
0 & 0 & 0 & -3
\end{array}\right)\ .
\end{align}
Note that $T_1,...,T_8$ generate an $SU(3)$ subgroup. 

The metric (\ref{gdef}) arises from \eqref{adsmetric} by setting $d\phi=0$ and taking $\phi\rightarrow\infty$. 
To obtain the symmetries of the boundary theory we must therefore drop the  $\partial_{\phi}$ components. 
Furthermore since we reduce along the $x^+$ direction to obtain the field theory $S_\Omega$, we must also drop $\partial_{+}$ components.  
We can then later reintroduce  new $\partial_{+}$ components to obtain conformal Killing vectors of the boundary metric (\ref{gdef}). 
It is not difficult
to verify that these Killing vectors generate an $SU(3,1)$
algebra via their Lie derivatives (indeed, this is guaranteed by construction). 

By taking appropriate linear combinations we can obtain
the conformal Killing vectors listed in section \ref{Setup}. In particular we
find four translations along $x^{i}$:
\begin{align}
\frac{1}{4}(K_{12}-K_{2}),\,\,\,\frac{1}{4}(K_{14}-K_{5}),\,\,\,\frac{1}{4}\left(K_{1}+K_{11}\right),\,\,\,\frac{1}{4}\left(K_{4}+K_{13}\right)\ ,
\end{align}
a translation along $x^{-}$:
\begin{align}
-\frac{1}{24}(3K_{3}+\sqrt{3}K_{8}+6K_{9}+2\sqrt{6}K_{15})\ ,
\end{align}
four rotations preserving the $\Omega$-tensor:
\begin{align}
\frac{1}{2}K_{6},\,\,\,\frac{1}{2}K_{7},\,\,\,\frac{1}{2\sqrt{3}}\left(K_{8}-\frac{1}{\sqrt{8}}K_{15}\right),\,\,\,\frac{1}{4}\left(-K_{3}+\frac{1}{\sqrt{3}}K_{8}+\frac{1}{\sqrt{6}}K_{15}\right)\ ,
\end{align}
a dilatation: 
\begin{align}
-\frac{1}{4}K_{10}\ ,
\end{align}
four type $VI$  conformal symmetries:
\begin{align}
\frac{1}{4}(K_{1}-K_{11}),\,\,\,\frac{1}{4}\left(K_{4}-K_{13}\right),\,\,\,\frac{1}{4}\left(K_{2}+K_{12}\right),\,\,\,\frac{1}{4}\left(K_{5}+K_{14}\right)\ , 
\end{align}
and the  type $VII$ conformal symmetry:
\begin{align}
-\frac{1}{48}(3K_{3}+\sqrt{3}K_{8}-6K_{9}+2\sqrt{6}K_{15})\ .	
\end{align}


\begin{thebibliography}{3}

\bibitem{Aharony:2008ug}
  O.~Aharony, O.~Bergman, D.~L.~Jafferis and J.~Maldacena,
  ``N=6 superconformal Chern-Simons-matter theories, M2-branes and their gravity duals,''
  JHEP {\bf 0810} (2008) 091
  doi:10.1088/1126-6708/2008/10/091
  [arXiv:0806.1218 [hep-th]].

\bibitem{Aharony:1997an}
  O.~Aharony, M.~Berkooz and N.~Seiberg,
  ``Light cone description of (2,0) superconformal theories in six-dimensions,''
  Adv.\ Theor.\ Math.\ Phys.\  {\bf 2} (1998) 119
  doi:10.4310/ATMP.1998.v2.n1.a5
  [hep-th/9712117].

\bibitem{Douglas:2010iu} 
  M.~R.~Douglas,
  ``On D=5 super Yang-Mills theory and (2,0) theory,''
  JHEP {\bf 1102}, 011 (2011)
  doi:10.1007/JHEP02(2011)011
  [arXiv:1012.2880 [hep-th]].

\bibitem{Lambert:2010iw} 
  N.~Lambert, C.~Papageorgakis and M.~Schmidt-Sommerfeld,
  ``M5-Branes, D4-Branes and Quantum five-dimensional super-Yang-Mills,''
  JHEP {\bf 1101}, 083 (2011)
  doi:10.1007/JHEP01(2011)083
  [arXiv:1012.2882 [hep-th]].
 
\bibitem{Hull:2014cxa} 
  C.~M.~Hull and N.~Lambert,
  ``Emergent Time and the M5-Brane,''
  JHEP {\bf 1406}, 016 (2014)
  doi:10.1007/JHEP06(2014)016
  [arXiv:1403.4532 [hep-th]].

\bibitem{Lambert:2011gb} 
  N.~Lambert and P.~Richmond,
  ``(2,0) Supersymmetry and the Light-Cone Description of M5-branes,''
  JHEP {\bf 1202}, 013 (2012)
  doi:10.1007/JHEP02(2012)013
  [arXiv:1109.6454 [hep-th]].

\bibitem{Lambert:2018lgt}
  N.~Lambert and M.~Owen,
  ``Non-Lorentzian Field Theories with Maximal Supersymmetry and Moduli Space Dynamics,''
  JHEP {\bf 1810} (2018) 133
  doi:10.1007/JHEP10(2018)133
  [arXiv:1808.02948 [hep-th]].

\bibitem{Mouland:2019zjr} 
  R.~Mouland,
  ``Supersymmetric Soliton $\sigma$-models from Non-Lorentzian Field Theories,''
  arXiv:1911.11504 [hep-th].

\bibitem{Lambert:2019nti} 
  N.~Lambert and R.~Mouland,
  ``Non-Lorentzian RG flows and Supersymmetry,''
  JHEP {\bf 1906}, 130 (2019)
  doi:10.1007/JHEP06(2019)130
  [arXiv:1904.05071 [hep-th]].

\bibitem{Lambert:2019jwi}
  N.~Lambert, A.~Lipstein and P.~Richmond,
  ``Non-Lorentzian M5-brane Theories from Holography,''
  JHEP {\bf 1908} (2019) 060
  doi:10.1007/JHEP08(2019)060
  [arXiv:1904.07547 [hep-th]].

\bibitem{Pope:1999xg} 
  C.~N.~Pope, A.~Sadrzadeh and S.~R.~Scuro,
  ``Timelike Hopf duality and type IIA* string solutions,''
  Class.\ Quant.\ Grav.\  {\bf 17}, 623 (2000)
  doi:10.1088/0264-9381/17/3/305
  [hep-th/9905161].

\bibitem{Beisert:2010jr} 
  N.~Beisert {\it et al.},
  ``Review of AdS/CFT Integrability: An Overview,''
  Lett.\ Math.\ Phys.\  {\bf 99}, 3 (2012)
  doi:10.1007/s11005-011-0529-2
  [arXiv:1012.3982 [hep-th]].

\bibitem{Beisert:2018zxs} 
  N.~Beisert, A.~Garus and M.~Rosso,
  ``Yangian Symmetry for the Action of Planar $\mathcal N=$ 4 Super Yang-Mills and $\mathcal N=$ 6 Super Chern-Simons Theories,''
  Phys.\ Rev.\ D {\bf 98}, no. 4, 046006 (2018)
  doi:10.1103/PhysRevD.98.046006
  [arXiv:1803.06310 [hep-th]].

\bibitem{Beem:2013sza} 
  C.~Beem, M.~Lemos, P.~Liendo, W.~Peelaers, L.~Rastelli and B.~C.~van Rees,
  ``Infinite Chiral Symmetry in Four Dimensions,''
  Commun.\ Math.\ Phys.\  {\bf 336}, no. 3, 1359 (2015)
  doi:10.1007/s00220-014-2272-x
  [arXiv:1312.5344 [hep-th]].
  
  \bibitem{Hoxha:2000jf} 
  P.~Hoxha, R.~R.~Martinez-Acosta and C.~N.~Pope,
  ``Kaluza-Klein consistency, Killing vectors, and Kahler spaces,''
  Class.\ Quant.\ Grav.\  {\bf 17}, 4207 (2000)
  doi:10.1088/0264-9381/17/20/305
  [hep-th/0005172].

\bibitem{Huang:2010qy} 
  Y.~t.~Huang and A.~E.~Lipstein,
  ``Dual Superconformal Symmetry of N=6 Chern-Simons Theory,''
  JHEP {\bf 1011}, 076 (2010)
  doi:10.1007/JHEP11(2010)076
  [arXiv:1008.0041 [hep-th]].



\end{thebibliography}
\end{document}